\documentstyle[12pt,epsf]{article}
\def\s{\sigma}
\def\half{{\textstyle{{1}\over{2}}}}

\def\nn{\nonumber}
\def\df{\partial}
\def\frac#1#2{{\textstyle{{#1}\over{#2}}}}
\newlength{\mywidth}\mywidth=2.75truein 
\renewenvironment{figure}{\refstepcounter{figure}
\baselineskip=0.4\normalbaselineskip\footnotesize}
{\baselineskip=\normalbaselineskip}
\def\fignum{{\bf Fig.\arabic{figure}.\quad}}
\def\fref#1{fig.\ref{#1}}

\def\frac#1#2{{\textstyle{{#1}\over{#2}}}}

\begin{document}

\title{String theory in Lorentz-invariant time-like gauge}
\author{Igor Nikitin\\
{\it GMD/IMK, 53754 St.Augustin, Germany} \\
E-mail: Igor.Nikitin@gmd.de
}
\date{}
\maketitle


\vspace{-5mm}
\begin{abstract}
A theory of closed bosonic string in time-like gauge, related in Lorentz-invariant way
with the world sheet, is considered. Absence of quantum anomalies in this theory is shown. 

\vspace{2mm}\noindent
PACS~11.25: theory of fundamental strings, noncritical string theory.

\end{abstract}

\vspace{-3mm}
\section*{Introduction}
Relativistic string is a curve moving in $d$-dimensional Minkowski space
and sweeping by its motion a 2-dimensional surface (world sheet).
An area of the world sheet is proposed to be an action of the string.

There are two approaches for the description of string dynamics.
{\it Covariant approach} keeps the main symmetry of string's action:
group of arbitrary reparametrizations of the world sheet, which should be implemented 
both in classical and quantum theories. {\it Non-covariant approach} eliminates this symmetry,
introducing a particular parametrization (gauge) on the world sheet.

In covariant approach the string theory is usually considered in oscillator representation,
analogous to the description of field theories by operators of creation and annihilation.
It is well known \cite{Brink}, that reparametrization group in this approach has anomalies, 
violating the parametrical invariance of quantum theory. In special case 
$d=26$ the quantum theory has peculiar properties (an existence of properly factorizable {\it null subspace}),
this is usually considered as an indication of its parametrical invariance.

Covariant string theory was also investigated in non-oscillator representations.
Particularly, a consideration, done by Mezincescu and Hennaux in \cite{Brink} p.157,
shows that the theory of closed bosonic string at arbitrary even number of dimensions,
quantized in $x,p$-representation, has a solution, possessing quantum parametrical invariance.
This fact indicates the absence of anomalies in such representation. Analogous result
was obtained in work \cite{indef}, which  shows, that quantum theory of open string
in pseudo-Euclidean space with equal number of spatial and temporal directions (particularly,
$d=3+3$) can be realized in positively defined extended space of states without anomalies
in group of reparametrizations.

Consideration of string theory in covariant approach is related with one difficulty.
It is shown in a recent work \cite{exotic} that the phase space of covariant string
theory is much wider than it is usually assumed. It contains infinite regions, filled 
by classical solutions
with negative energies and negative square of mass, related with existence of
folds on the world sheets, see \fref{f1}. Conditions, excluding such solutions in classical mechanics,
cannot be directly transferred to covariant quantum string 
theory\footnote{Work \cite{exotic} shows that standard covariant quantization of string 
theory in oscillator representation does not exclude classical tachionic solutions,
but due to a special effect, caused by indefiniteness of space 
of states, {\it redefines} their square of mass in such a way that $M^{2}<0\ \to\ M^{2}>0$, 
and then {\it identifies} the tachionic solutions with normal ones.}. 
Consistent exclusion of tachionic solutions is possible only in
non-covariant approach, when one chooses a special parametrization of the world sheet.

\begin{center}
\parbox{5cm}{\begin{figure}\label{f1}
\begin{center}
~\epsfysize=5cm\epsfxsize=5cm\epsffile{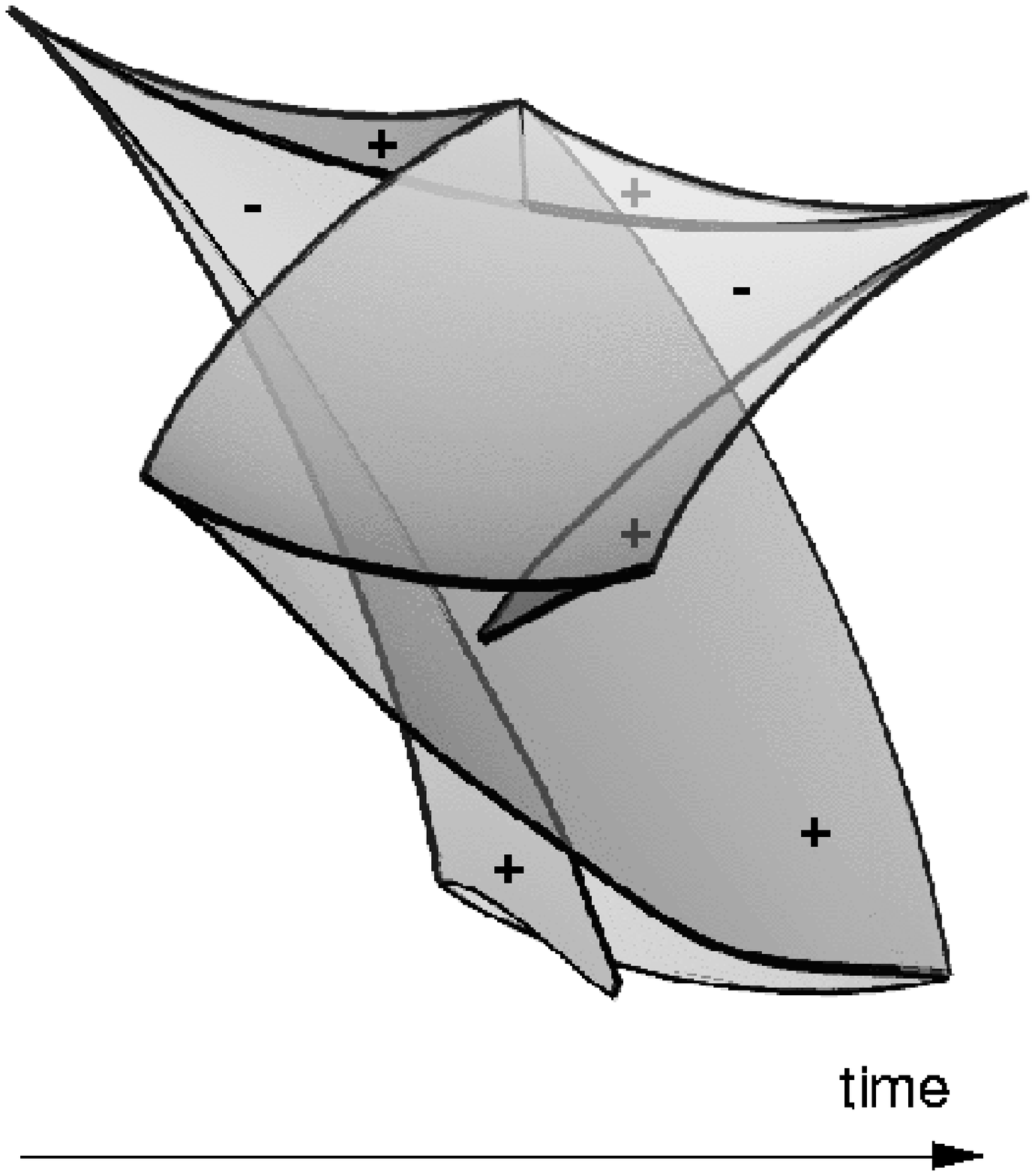}
\end{center}

\fignum Fragment of the world sheet for negative energy solution.
Computer generated image from \cite{exotic}.

\end{figure}
}\quad\quad
\parbox{5cm}{\begin{figure}\label{f2}
\begin{center}
~\epsfysize=5cm\epsfxsize=5cm\epsffile{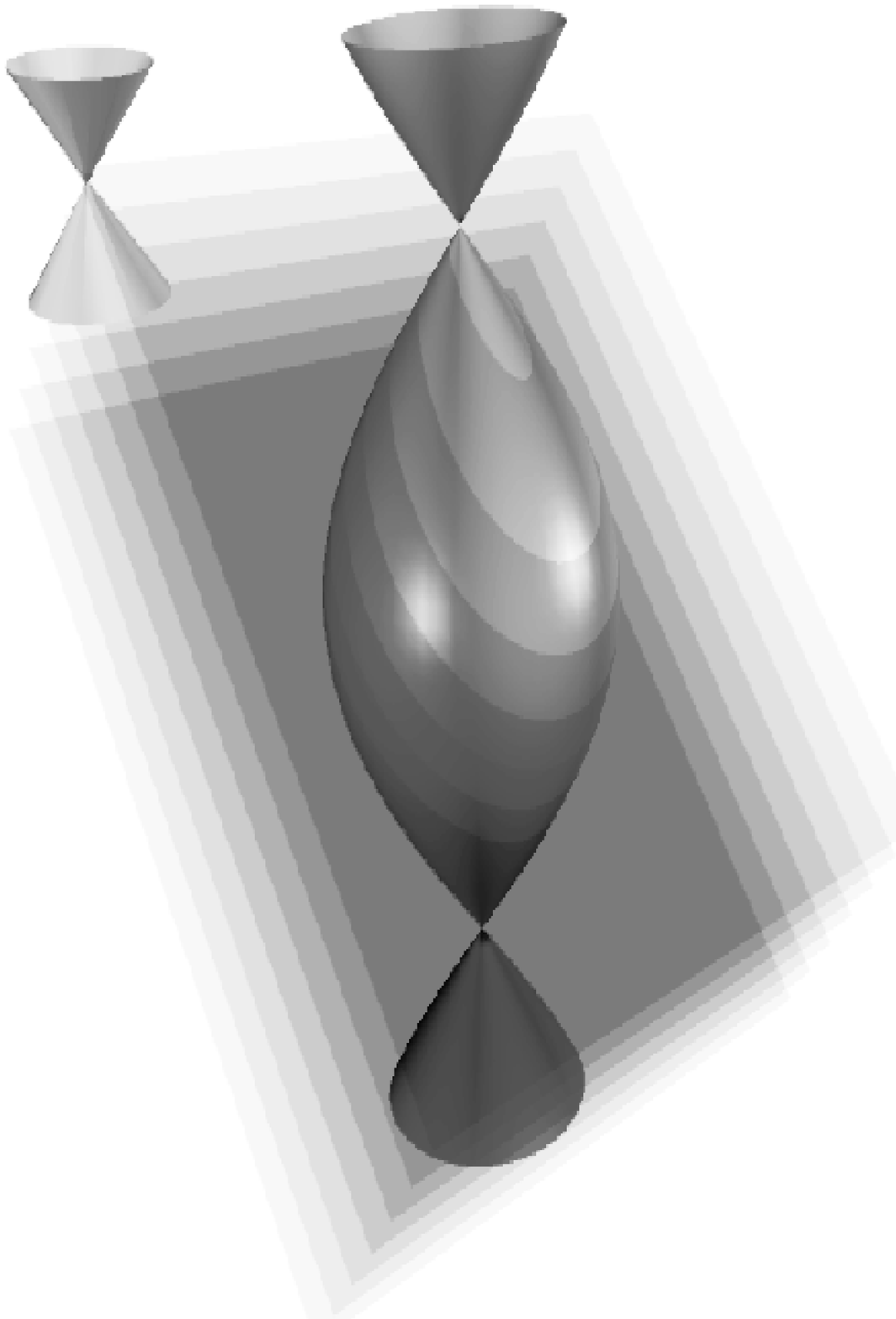}
\end{center}

\fignum Light-like parametrization on the world sheet. 

\end{figure}
}\quad\quad
\parbox{5cm}{\begin{figure}\label{f3}
\begin{center}
~\epsfysize=5cm\epsfxsize=5cm\epsffile{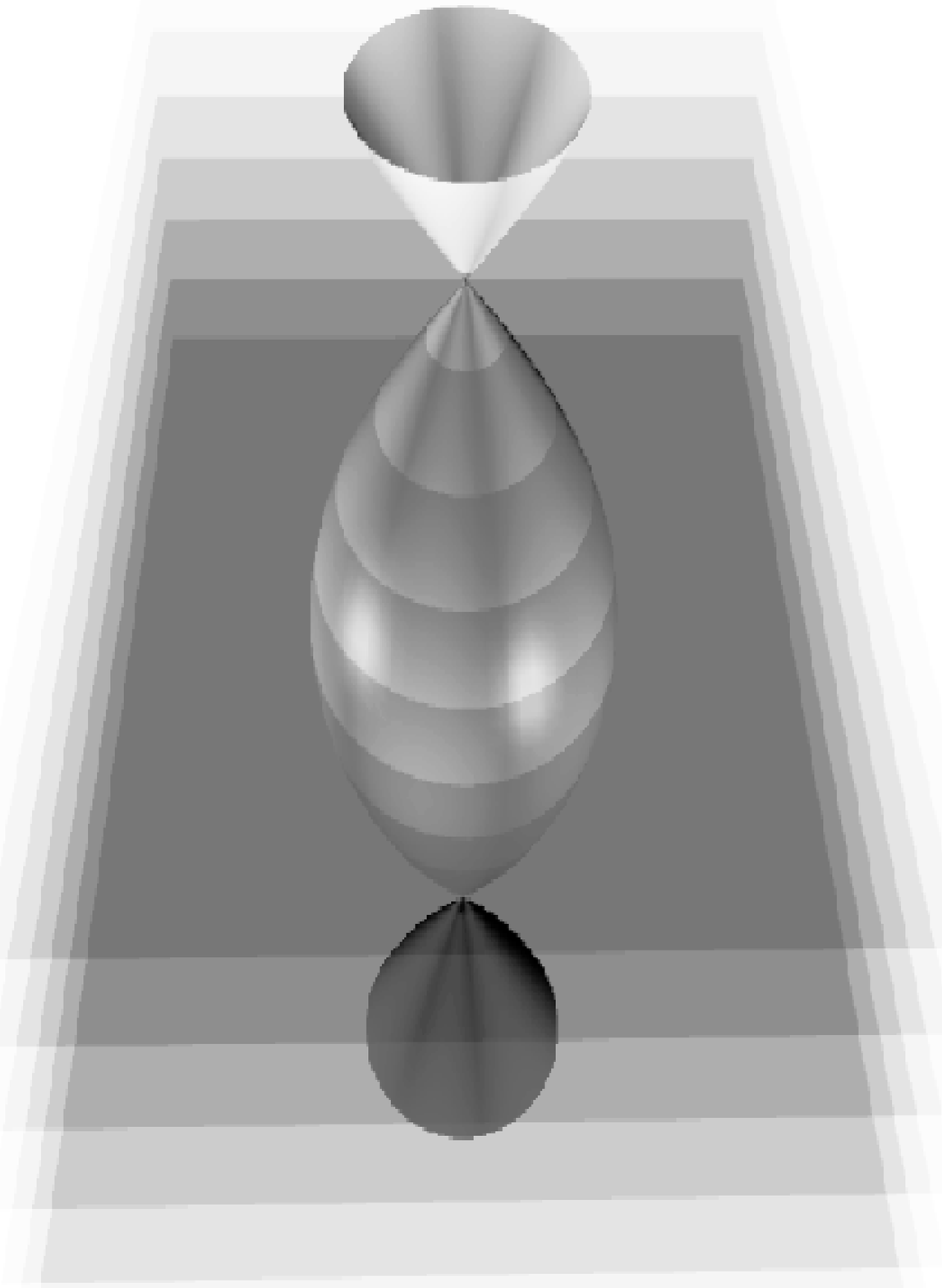}
\end{center}

\fignum Time-like parametrization on the world sheet. 

\end{figure}
}
\end{center}

\vspace{7mm}
Standard parametrization, used in non-covariant approach is {\it light cone gauge}.
This parametrization can be obtained in slicing of the world sheet by a set of parallel planes,
where one is tangent to light cone with a center in the origin (\fref{f2}). 
Usually the position of such planes is ``frozen'' in the space-time, i.e. Lorentz transformations change 
the location of the world sheet with respect to the slicing planes. As a result, Lorentz transformations 
change the slices on the world sheet, i.e. are followed by reparametrization.
In this way the common anomaly in reparametrization group appears also in the group
of Lorentz transformations\footnote{Exception is a case $d=26$, where the anomalies are absent.}.

A simplest way to avoid this problem was proposed in \cite{ax,dlcg}. In these works 
{\it Lorentz-invariant} light cone gauge was introduced, which relates
the slicing planes with the world sheet itself, so that Lorentz transformations
move them together the world sheet. In this approach Lorentz transformations are not
followed by reparametrization, and on quantum level have no anomalies. Quantum mechanics,
based on this idea, was studied by different approaches in \cite{dlcg}. 

Another Lorentz-invariant parametrization: time-like gauge in center-of-mass frame was introduced
by Rohrlich \cite{Rohrlich}.
This gauge leads to a complicated Hamiltonian mechanics.
In full generality this mechanics was investigated in \cite{zone}.
In \cite{slstring,2par} particular finite-dimensional subsets in the phase space were selected,
admitting anomaly-free quantization in $d=3+1$.

In the present work we describe another Lorentz-invariant gauge, which can be considered
as weakened variant of the Rohrlich's one. This gauge leads to a simple
Hamiltonian mechanics and gives a possibility to construct anomaly-free quantum theory
at arbitrary number of dimensions.

\section{Classical mechanics}
Theory of closed bosonic string in $d$-dimensional Minkowski space-time is described 
by canonically conjugated coordinates and momenta:
\begin{eqnarray}
&&\{x_{\mu}(\s),p_{\nu}(\tilde\s)\}=g_{\mu\nu}\Delta(\s-\tilde\s),\quad \mu,\nu=0...d-1.\label{var0}
\end{eqnarray}
Here $x_{\mu}(\s),p_{\mu}(\s)$ are $2\pi$-periodical functions and $\Delta(\s)$ is $2\pi$-periodical
Dirac's delta-function. Coordinates and momenta are restricted by constraints:
\begin{eqnarray}
&&x'p=0,\quad x'^{2}+p^{2}=0.\label{con0}
\end{eqnarray}
The constraints belong to the first class in Dirac's terminology \cite{Dirac}: 
Poisson brackets of constraints vanish on their surface.

\paragraph*{Mechanics in center-of-mass frame.} 
Total momentum of the string is given by expression $P_{\mu}=\oint d\s p_{\mu}(\s)$
(here and further $\oint d\s$ denotes $\int_{0}^{2\pi}d\s$).
Following \cite{zone}, introduce orthonormal 
basis of vectors, dependent on total momentum: $N^{\alpha}_{\mu}(P),\
N^{\alpha}_{\mu}N^{\beta}_{\mu}=g^{\alpha\beta}={\rm diag}(+1,-1,..,-1)$, with
$N^{0}_{\mu}=P_{\mu}/\sqrt{P^{2}}$. This basis defines center-of-mass frame (CMF),
where $N^{0}_{\mu}$ is temporal axis (directed along total momentum $P_{\mu}$)
and $N^{i}_{\mu},\ i=1..d-1$ are spatial axes (orthogonal to $P_{\mu}$). 

String dynamics in CMF is defined by a set of new canonical variables 
$(Z_{\mu},P_{\mu},q^{\alpha}(\s),p^{\alpha}(\s))$ with Poisson brackets
\begin{eqnarray}
&&\{Z_{\mu},P_{\nu}\}=g_{\mu\nu},\quad \{Z_{\mu},p^{0}(\s)\}=N^{0}_{\mu}\Delta(\s),\quad
\{q^{\alpha}(\s),p^{\beta}(\tilde\s)\}=g^{\alpha\beta}(\Delta(\s-\tilde\s)-\Delta(\tilde\s)),\label{var1}
\end{eqnarray}
other Poisson brackets vanish. New variables are related with old ones by expressions
(for detailed proofs and derivations look to Appendix~1): 
$$q^{\alpha}(\s)=x^{\alpha}(\s)-x^{\alpha}(0),\quad 
x^{\alpha}(\s)=N^{\alpha}_{\mu}x_{\mu}(\s),\quad
p^{\alpha}(\s)=N^{\alpha}_{\mu}p_{\mu}(\s),$$
$x^{\alpha},p^{\alpha}$ are projections of coordinates and momenta onto axes of CMF;
mean coordinate $Z_{\mu}$:
$$Z_{\mu}=N^{0}_{\mu}x^{0}(0)+\half N^{\alpha}_{\nu}(\df N^{\beta}_{\nu}/\df P_{\mu})
M^{\alpha\beta},\quad M^{\alpha\beta}=\oint d\s x^{[\alpha}p^{\beta]}.$$
Here square brackets denote antisymmetrization: $x^{[\alpha}p^{\beta]} =x^{\alpha}p^{\beta}-
x^{\beta}p^{\alpha} $. Variables $M^{\alpha\beta}$ can be obtained from usual Lorentz generators 
$M_{\mu\nu}=\oint d\s x_{[\mu}p_{\nu]}$ by projection to CMF. 

New canonical variables are restricted by constraints:
\begin{eqnarray}
q^{\alpha} (0)=0,\quad \delta^{\alpha0}\sqrt{P^{2}}-\oint d\s p^{\alpha}(\s)=0,\quad
q'^{0}p^{0}-\vec q{\;}'\vec p=0,\quad (q'^{0})^{2}+(p^{0})^{2}-\vec q{\;}'^{2}-\vec p{\;}^{2}=0,&&\label{con1}
\end{eqnarray}
which as earlier belong to the first class. 

\paragraph*{Time-like gauge in CMF.} Let's impose a condition (gauge): $q^{0}(\s)\equiv0$. 
It introduces a particular parametrization on the world sheet, requiring that
the string $x_{\mu}(\s)$ should always be an equal-time slice of the world sheet in CMF, see \fref{f3}.
This gauge gives a possibility to exclude $p^{0}(\s)$ from the set of independent variables,
defining it identically as $p^{0}\equiv\sqrt{\vec q{\;}'^{2}+\vec p{\;}^{2}}$. Remaining variables
have Poisson brackets:
\begin{eqnarray}
&&\{Z_{\mu},P_{\nu}\}=g_{\mu\nu},\quad 
\{q^{i}(\s),p^{j}(\tilde\s)\}=-\delta^{ij}(\Delta(\s-\tilde\s)-\Delta(\tilde\s)),\ i,j=1..d-1\label{var2}
\end{eqnarray}
(others are zero). The constraints
\begin{eqnarray}
&&\vec q (0)=0,\quad\oint d\s \vec p(\s)=0,\quad
\vec q{\;}'\vec p=0,\quad \sqrt{P^{2}}-\oint d\s\sqrt{\vec q{\;}'^{2}+\vec p{\;}^{2}}=0\label{con2}
\end{eqnarray}
are again of the first class. In the spirit of Dirac's theory of constrained systems \cite{Dirac},
each constraint, being used as Hamiltonian, generates canonical transformations
in the phase space, representing symmetries of the system.  In our case the constraints
generate reparametrizations of the world sheet, i.e. the transformations, which preserve the
action.

\noindent
\parbox{2.9cm}{\begin{figure}\label{f4}
\begin{center}
~\epsfysize=2cm\epsfxsize=2cm\epsffile{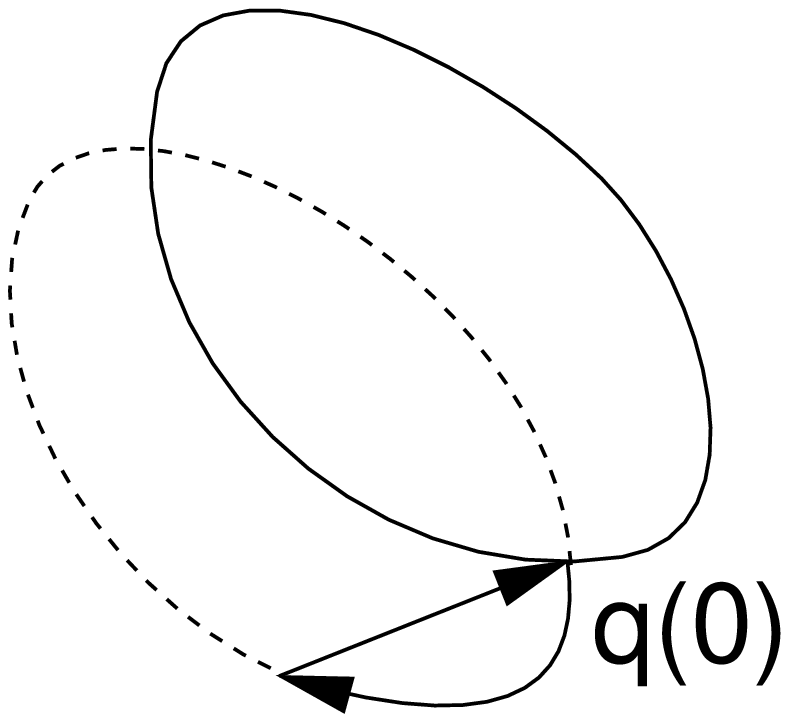}
\end{center}
\fignum~The~third~constraint~generates reparametrizations and~translations.
\end{figure}
}\quad\quad
\parbox{14cm}{In more details: from (\ref{var2}) one can conclude that the first two constraints in
(\ref{con2}) have identically vanishing Poisson brackets with all dynamical variables 
and generate no transformations (they are auxiliary elements of our construction,
not related with any symmetry of the system). The third constraint generates reparametrizations
of the string $q^{i}(\s)\to q^{i}(\tilde\s)$, followed by a translation, shifting new $q^{i}(\tilde\s=0)$
to the origin -- see \fref{f4}. This translation is necessary due to the first constraint in (\ref{con2}).
The same translation, but in opposite direction, is applied to $x^{i}(0)$ (see Appendix~1),
resulting to pure reparametrization of equal-time slice $x^{i}(\s)$. The fourth constraint generates
translations of slicing plane in $P_{\mu}$ direction, and correspondent evolution of 
equal-time slice, generated by Hamiltonian $H=\oint d\s\sqrt{\vec q{\;}'^{2}+\vec p{\;}^{2}}$.
This evolution has a period $T=\sqrt{P^{2}}$ (evolution, generated by an equivalent constraint
$(P^{2}-H^{2})/2\pi=0$, is $2\pi$-periodic).
}

\paragraph*{Lorentz generators}
are given by expressions, identical to \cite{slstring}:
\begin{eqnarray}
&&M_{\mu\nu}=X_{[\mu}P_{\nu]}+N_{\mu}^{i}N_{\nu}^{j}M^{ij},\quad
X_{\mu}=Z_{\mu}- \half N^{i}_{\nu}(\df N^{j}_{\nu}/\df P_{\mu})M^{ij}.\label{Lor}
\end{eqnarray}
Here $M^{ij}$ is the tensor of orbital moment of the string in CMF, which in bosonic string model 
is identified with the spin of the particle. In the case $d=3+1$ it is related with spin vector 
$\vec M=\oint d\s\vec x\times\vec p$ as $M^{ij}=\epsilon^{ijk}M^{k}.$ 

\vspace{2mm}\noindent
{\it The clue property:} if in quantum mechanics the commutators of $Z_{\mu},P_{\mu}$
are postulated directly from Poisson brackets (\ref{var2}), and the algebra of CMF-rotations $SO(d-1)$
is represented correctly by $M^{ij}$: e.g. $[M^{i},M^{j}]=i\epsilon_{ijk}M^{k}$ for $d=4$,
then operators, defined by expressions (\ref{Lor}), represent correctly the algebra of Lorentz
transformations $SO(d-1,1)$. For $d=4$ this was proven in \cite{slstring} by direct calculation, 
and for other $d$ can be proven analogously. Natural explanation of this fact was also given
in \cite{dlcg}: variables (\ref{Lor}) actually generate Lorentz transformations of the world
sheet together with the set of slicing planes. They are not followed by
reparametrizations of the world sheet, like in standard light-cone gauge. Namely these auxiliary
reparametrizations create problems in standard approach, because they bring anomalies, destroying 
the Lorentz algebra.

\section{Quantum mechanics}\label{S2}
Canonical commutators
\begin{eqnarray}
&&[Z_{\mu},P_{\nu}]=-ig_{\mu\nu},\quad 
[q^{i}(\s),p^{j}(\tilde\s)]=i\delta^{ij}(\Delta(\s-\tilde\s)-\Delta(\tilde\s))
\nn
\end{eqnarray}
can be realized in a direct product of space of functions $\phi(P)$ onto the space of
functionals $\psi[q(\s)]$, with definition of operators 
\begin{eqnarray}
&&Z_{\mu}=-i{{\df}\over{\df P_{\mu}}},\quad
p^{i}(\s)=-i\left( {{\delta}\over{\delta q^{i}(\s)}} -\Delta(\s)\cdot\oint d\tilde\s
{{\delta}\over{\delta q^{i}(\tilde\s)}}\right).\label{varq3}
\end{eqnarray}
Constraints:
\begin{eqnarray}
\oint d\s \vec p(\s)\;\psi =0; \quad \vec q (0)\;\psi =0;\quad 
\vec q{\;}'\vec p\;\psi =0;\quad (\sqrt{P^{2}}-H)\psi=0,\quad H= \oint d\s\sqrt{\vec q{\;}'^{2}+\vec p{\;}^{2}}.&&
\label{conq2}
\end{eqnarray}
From the definition of $\vec p$ we see that the first constraint is satisfied identically for any $\psi$.
To satisfy the second constraint, we should define $\psi=\delta(\vec q(0))\Psi$. Then, we see that
$\vec p$ commutes through $\delta$-factor: $\vec p\delta(\vec q(0))\Psi=\delta(\vec q(0))\vec p\;\Psi$.
The second term in $\vec p$ is generator of global translations $\vec q(\s)\to\vec q(\s)+\vec \epsilon$.
Requiring that $\Psi$ is translationally invariant\footnote{In the presence of $\delta(\vec q(0))$
the values $\Psi[\vec q(\s)]$ for loops with $\vec q(0)\neq0$ are not important. Assuming translational
invariance of $\Psi$, we actually define these values to be the same as for loop with $\vec q(0)=0$. 
This assumption simplifies further calculations.}:  $\Psi[\vec q(\s)+\vec\epsilon\;]=\Psi[\vec q(\s)]$,
we will have $\delta(\vec q(0))\vec p\;\Psi=\delta(\vec q(0))(-i \delta/\delta\vec q(\s) )\Psi$.
For the third constraint: considering linear combinations $\oint d\s\epsilon(\s)\vec q{\;}'(\s)
\delta/\delta\vec q(\s) $, we see that they act on the state $\Psi$ as generators of reparametrizations
$\vec q(\s)\to\vec q(\s)+\epsilon(\s)\vec q{\;}'(\s)$. Requiring additionally 
that $\Psi$ is parametrical invariant: $\Psi[\vec q(\tilde\s)]=\Psi[\vec q(\s)]$,
we will satisfy the third constraint. The fourth constraint has a form of mass shell condition.
To satisfy it, we should solve eigenvalue problem for operator $H$ (this will automatically
determine the spectrum of mass).

\paragraph*{Operator H} can be defined using an expansion of the square root:
\begin{eqnarray}
&&H=\oint d\s \sqrt{\vec q{\;}'^{2}}\left(
1+{{1}\over{2\vec q{\;}'^{2}}}\vec p{\;}^{2}-{{1}\over{(4\vec q{\;}'^{2})^{2}}}
(\vec p{\;}^{2})^{2}+...
\right).\label{H}
\end{eqnarray}
Our next goal is to show, that each term of this expansion acts in the selected space of states,
i.e. functions of the form $\delta(\vec q(0))\Psi$, where $\Psi$ are translationally and 
parametrically invariant functionals of $\vec q(\s)$.

Formal calculation, given in Appendix~2, shows that each term in (\ref{H}) commutes with the first
three constraints in (\ref{conq2}), and therefore should act in the selected space of states.
However, in concrete calculations the definition of $H$ should be refined, because powers 
of variational derivative $\delta/\delta q^{i}$ create divergencies. 
Particularly,  action of  $\delta/\delta q^{i}(\s)$
on parametrically invariant integrals $\oint d\s q'^{j}F^{j}(q)$ gives local expression
$q'^{j}\df F^{[j}/\df q^{i]}|_{\s}$, and (in the case if $\df F^{[j}/\df q^{i]}$ is not constant)
the next differentiation $\delta/\delta q^{i}(\s)$ gives a divergency $\Delta(0)\cdot (q'^{j}
\df^{2} F^{[j}/\df q^{i]}\df q^{i})$. Further we will introduce a wide set of translationally 
and parametrically invariant functionals, for which $H$ can be reasonably defined.

Let's consider parametrically invariant functionals of the form $A^{i}(k)=\oint d\s q'^{i}e^{ikq}$, 
and took their translationally invariant products: $1,\ A^{i}(k)A^{j}(-k),\ 
...,\
A^{i_{1}}(k_{1})..A^{i_{n}}(k_{n})\delta(k_{1}+...+k_{n})$. Let's define the powers
of variational derivative as $\lim\limits_{\tilde\s\to\s}\delta/\delta q^{i}(\tilde\s)\cdot\delta/\delta q^{i}(\s)$
(classically this corresponds to the same variable $\lim\limits_{\tilde\s\to\s}p^{i}(\tilde\s)p^{i}(\s)=
p^{i}(\s)^{2}$).
In this definition the divergent terms are omitted (each $A^{i}(k)$ in the product is differentiated
once, giving local expression of $\s$, and further variational derivatives with respect to
$\vec q(\tilde\s\neq\s)$ are not applied to it).
Note that for each product finitely many variational derivatives can be applied to give non-vanishing
result, and expansion (\ref{H}) is actually truncated to a finite sum. After all differentiations
and outer integration $\oint d\s$ in (\ref{H}) we obtain well-defined functional. Particularly,
\begin{eqnarray}
&&H\ A^{i}(k)A^{i}(-k)=\oint d\s\sqrt{q'^{2}}\cdot A^{i}(k)A^{i}(-k)-
\oint d\s\sqrt{q'^{2}}\left(k^{2}+(d-2){{(q'k)^{2}}\over{q'^{2}}}\right).\nn
\end{eqnarray}
Analogous expressions can be written for other products.

\paragraph*{Remarks}\ 

\vspace{1mm}
\noindent 1. The functionals, found in action of $H$ to the products of $A^{i}(k)$, are
translationally and parametrically invariant, however, they are not expressed as products of $A^{i}(k)$.
If one finds the expansion of resulting functionals by the products of $A^{i}(k)$,
i.e. prove that $H$ acts in the space, spanned by these products, 
this will give a possibility to solve eigenvalue problem for $H$.

\vspace{1mm}
\noindent 2. Another technically difficult task is an introduction of Hermitian structure 
in this space of states. It is known \cite{sp}, that in theories with non-compact gauge
group the scalar product from extended space of states is divergent on physical subspace,
selected by constraints (it includes an infinite volume of gauge group). Therefore, it is needed
to introduce a new scalar product, acting in the
physical subspace. Hermitian property is required only for operators,
acting in the physical subspace, particularly, each term in expansion (\ref{H}),
integrated by $\oint d\s$, should be Hermitian operator (not $q'(\s),p(\s)$, from which
it is composed -- separately these operators do not act in the physical subspace).

\vspace{1mm}
\noindent 3. It was noted in the previous section, that the function  $H^{2}/2\pi$
is action-type variable, generating $2\pi$-periodical evolution. Quasiclassically this means
that $H^{2}/2\pi$ takes integer eigenvalues ($e^{2\pi i(H^{2}/2\pi)}=1$). This property also reflects
a general symmetry of mechanics \cite{sing}: translation by $P_{\mu}$ transforms the world sheet to itself.
(However, concrete definitions of $H$ can violate this property, and the described symmetry can be 
lost.)

\vspace{1mm}
\noindent 4. Lorentz group in our approach is free of anomalies, because in generator of
CMF-rotations $\vec M=\oint d\s\vec q\times\vec p=-i\oint d\s\vec q\times\delta/\delta\vec q
+i\vec q(0)\times\oint d\s\delta/\delta\vec q$ the second term vanishes on translationally invariant
functionals (also due to the presence of $\delta(\vec q(0))$ in $\psi$), and the first term defines
correct representation of rotation group in considered space of states (acting as rotations of
the argument in $\Psi[\vec q(\s)]$). Rotations obviously commute with all constraints and Hamiltonian
$H$, therefore, the eigenspaces of $H$ should be rotationally invariant, and can be decomposed
into a sum of irreducible representations of rotation group, correspondent to definite values of spin.

\vspace{1mm}
\noindent 5. Absence of anomalies in this approach can be also understood in standard
oscillator representation of string theory. Here the constraint $\vec q{\;}'\vec p=0$ can be rewritten
as $L_{n}-\tilde L_{-n}=0$, where $L_{n},\tilde L_{n}$ are generators of Virasoro algebra
for left and right modes. Using a commutator $[L_{n},L_{m}]=(n-m)L_{n+m}+(d-1)/12\cdot
n(n^{2}-1)\delta_{n,-m}$ and the same commutator for $\tilde L_{n}$ (coefficient $(d-1)$ is here, 
because $L_{n},\tilde L_{n}$ include only $(d-1)$ space-like oscillators from CMF), one can see that
$L_{n}-\tilde L_{-n}$ forms closed Virasoro algebra without central charge.
Actually, we have imposed a gauge, eliminating anomalous components in these two Virasoro algebras,
and preserving their anomaly-free combinations.

\vspace{1mm}
\noindent 6. The gauge introduced here is quite similar to the time-like gauge proposed by Rohrlich 
\cite{Rohrlich}, used also in works [7-9], 
An important difference: the  Rohrlich's gauge,
except of $q^{0}=0$, also supposes $p^{0}=\sqrt{\vec q{\;}'^{2}+\vec p{\;}^{2}}=Const$. 
This additional constraint does not commute with $\vec q{\;}'\vec p=0$ 
(Rohrlich's gauge fixes parametrization on equal time slice --
our approach preserves this freedom). As a result, the whole set of constraints, appearing instead of
(\ref{con2}), is of the second class, and the reduction to these constraints leads to a complicated
Hamiltonian mechanics. 

There is one more distinctive feature of the approach \cite{Rohrlich} -- in this work the second class
constraints of the mechanics were imposed onto state vectors, in the spirit: 
two real constraints $x=0,\ p=0,\ [x,p]=i\ \to\ $ one complex constraint $(x+ip)\Psi=0$.
Such interpretation of constraints has some physical ground (explanation was given in \cite{dlcg}),
however, it is not equivalent to the standard interpretation \cite{Dirac}, where imposition
of second class constraints is possible only after reduction on their surface and quantization
of the obtained reduced mechanics. The present work and 
[7-9] 
use standard methods of constraints imposition.

\section*{Conclusion}
This work describes a particular gauge in the theory of closed bosonic string, 
which is applicable in the space-time with any number of dimensions,
and on the quantum level guarantees the absence of anomalies in Lorentz group and
the rest of reparametrization group. The main problem consists in a suitable definition
and determination of spectrum for a single operator: quantum analog of Hamiltonian
$H= \oint d\s\sqrt{\vec q{\;}'^{2}+\vec p{\;}^{2}}$.

\paragraph*{Acknowledgment.}
The work has been partially supported by \hbox{INTAS~96-0778} grant.

\baselineskip=0.4\normalbaselineskip\footnotesize

\section*{Appendix 1: canonical transformations}
\noindent 1. Canonical variables, defining string dynamics in CMF, were derived in \cite{zone},
using slightly different representation of string theory. Here we will reproduce this derivation
in $x,p$-representation. For this purpose we will use a formalism of symplectic forms 
\cite{slstring,Arnold}.

Poisson brackets correspond to a closed non-degenerate 
differential 2-form $\Omega = {{1}\over{2}}\omega_{ij}dX^{i}\wedge dX^{j}$, defined
on the phase space of the system (which is generally considered as a smooth manifold, 
endowed by local coordinates $X^{i},\  i=1,\ldots,2n$).
 Coefficient matrix of the form $\omega_{ij}$ is inverse to
the matrix of Poisson brackets $\omega^{ij}=\{X^{i},X^{j}\}$: 
$\omega_{ij}\omega^{jk}=\delta_{i}^{k}$.

Let's consider a surface in the phase space, given by the 2nd class 
constraints:
 $\chi_{\alpha}(X)=0\ (\alpha =1,\ldots,r),$\\ $det\|\{\chi_{\alpha},
\chi_{\beta}\}\|\neq 0 $. Reduction on this surface consists in the 
substitution of
its explicit parametrization $X^{i}=X^{i}(u^{a})\ (a=1,\ldots,2n-r)$ into the 
form: 
$$\Omega = \half\Omega_{ab}du^{a}\wedge du^{b},\quad
 \Omega_{ab}={{\partial X^{i}}\over{\partial u^{a}}}\omega_{ij}
 {{\partial X^{j}}\over{\partial u^{b}}},\quad
 det\|\Omega_{ab}\|\neq 0. $$
Matrix $\|\Omega^{ab}\|$ , inverse to 
$\|\Omega_{ab}\|$, defines Poisson brackets on the surface:
 $\{u^{a},u^{b}\}=\Omega^{ab}$.

This method is equivalent to commonly used Dirac brackets' formalism.
Sometimes it is convenient to combine both methods: some of the constraints
$\chi_{\alpha}(X)$ are imposed as above, then Dirac brackets on the remaining
constraints $\psi_{n}(u)$ are calculated by definition:
$$ \{u^{a},u^{b}\}^{D}=\{u^{a},u^{b}\} - \{u^{a},\psi_{n}\}\Pi^{nm}
\{\psi_{m},u^{b}\},$$
where $\|\Pi^{nm}\|$ is inverse to $\|\Pi_{nm}\|$: $\Pi_{nm}=
\{\psi_{n},\psi_{m}\}$.

In string theory canonical Poisson brackets 
$\{x_{\mu}(\s),p_{\nu}(\tilde\s)\}=g_{\mu\nu}\Delta(\s-\tilde\s)$
correspond to symplectic form $\Omega=\oint d\s\;
\delta p_{\mu}(\s)\wedge \delta x_{\mu}(\s)$, which is exact, i.e. can be represented as a
differential of 1-form: $\Omega=\delta\Psi,\ \Psi=\oint d\s\;p_{\mu}\delta x_{\mu}$.
Using decomposition $x_{\mu}=N^{\alpha}_{\mu}x^{\alpha},\ p_{\mu}=N^{\alpha}_{\mu}
p^{\alpha},$ we will have $\Psi=N^{\alpha}_{\mu}dN^{\beta}_{\mu}\oint d\s\;p^{\alpha}
x^{\beta}+\oint d\s\;p^{\alpha}\delta x^{\alpha}$, or after elementary transformations:
$\Psi=-\half N^{\alpha}_{\mu}(\df N^{\beta}_{\mu}/\df P_{\nu})M^{\alpha\beta}dP_{\nu}+
\oint d\s\;p^{\alpha}\delta x^{\alpha}$. Substituting $x^{\alpha}(\s)=x^{\alpha}(0)+q^{\alpha}(\s)$
and taking into account the identity $\oint d\s p^{i}=0$, we can rewrite the second term in 
$\Psi$ as $\sqrt{P^{2}}dx^{0}(0)+\oint d\s\;p^{\alpha}\delta q^{\alpha}$. Representing the first
term in this expression as $-x^{0}(0)(P_{\nu}/\sqrt{P^{2}})dP_{\nu}+d(\sqrt{P^{2}}x^{0}(0))$,
we will have $$\Psi=-Z_{\nu}dP_{\nu}+\oint d\s\;p^{\alpha}\delta q^{\alpha}+
\mbox{complete differential,\quad where\ } Z_{\nu}=x^{0}(0)(P_{\nu}/\sqrt{P^{2}})+
\half N^{\alpha}_{\mu}(\df N^{\beta}_{\mu}/\df P_{\nu})M^{\alpha\beta}.$$

We have constructed new variables in terms of old ones: $(x_{\mu},p_{\mu})\ \to\ 
(Z_{\mu},P_{\mu},q^{\alpha},p^{\alpha})$. It is necessary to show that old variables
can be reexpressed in terms of new ones, i.e. this mapping is invertible and we actually consider
two equivalent bases in the phase space. First of all, it is needed to reconstruct variable $x(0)$.
Considering $M^{\alpha\beta}$, we see that $x^{\alpha}(0)$ enters only in $M^{i0}$, and other
components are expressed completely in terms of new variables:
\begin{eqnarray}
&&M^{i0}=x^{i}(0)\sqrt{P^{2}}+\oint d\s q^{[i}p^{0]},\quad M^{ij}=\oint d\s q^{[i}p^{j]}.\label{Mi0}
\end{eqnarray}
Then we are able to obtain the required expression for $x(0)$:
\begin{eqnarray}
&&x_{\nu}(0)=Z_{\nu}+(P^{2})^{-1/2}N_{\nu}^{i}\oint d\s q^{[i}p^{0]}-
\half N^{i}_{\mu}(\df N^{j}_{\mu}/\df P_{\nu})M^{ij}.\label{x0}
\end{eqnarray}
Finally, old variables are reconstructed by relations 
$x_{\mu}(\s)=x_{\mu}(0)+N^{\alpha}_{\mu}q^{\alpha}(\s),\ p_{\mu}(\s)=N^{\alpha}_{\mu}
p^{\alpha}(\s).$ 

Now we can find symplectic form $\Omega$ (using antisymmetry of $\wedge$-operation and property
$d^{2}=0$):
\begin{eqnarray}
&&\Omega=d\Psi=dP_{\mu}\wedge dZ_{\mu}+\oint d\s\;\delta p^{\alpha}\wedge\delta q^{\alpha},
\label{Om0}
\end{eqnarray}
and inverting its coefficient matrix, find Poisson brackets:
\begin{eqnarray}
&&\{Z_{\mu},P_{\nu}\}=g_{\mu\nu},\quad 
\{q^{\alpha}(\s),p^{\beta}(\tilde\s)\}=g^{\alpha\beta}\Delta(\s-\tilde\s).\nn
\end{eqnarray}
Variables are restricted by constraints:
\begin{eqnarray}
&&q^{\alpha} (0)=0,\quad g^{\alpha0}\sqrt{P^{2}}-\oint d\s p^{\alpha}(\s)=0,\quad
q'p=0,\quad q'^{2}+p^{2}=0,\label{conA1}
\end{eqnarray}
The first two constraints have non-zero Poisson brackets, proportional to $g^{\alpha\beta}$,
and belong to the second class. Calculating new Poisson brackets (= Dirac's brackets on the surface
of second class constraints), we obtain expressions (\ref{var1}). After this procedure 
the first two constraints have vanishing Poisson brackets (are in involution)
with all dynamical variables\footnote{This is guaranteed by a structure of Dirac's brackets 
and also evident from definition  (\ref{var1}).}, and as a result, with all constraints. 
The Poisson brackets of derivatives $q'^{\alpha}$ and 
momenta $p^{\beta}$ have the same structure, as those for $x'_{\mu}$ and  $p_{\nu}$, 
therefore, the second pair of constraints (\ref{conA1}) obeys the same algebra as (\ref{con0}). 
Finally, the whole set of constraints belong to the first class.

\noindent 2. Imposing the gauge $q^{0}=0$ in symplectic form (\ref{Om0}), we see that conjugated
variable $p^{0}$ drops out. It can be expressed from the fourth constraint in (\ref{conA1})
as $p^{0}=\sqrt{\vec q{\;}'^{2}+\vec p{\;}^{2}}$. Calculating Poisson brackets for the obtained form,
we have
\begin{eqnarray}
&&\{Z_{\mu},P_{\nu}\}=g_{\mu\nu},\quad 
\{q^{i}(\s),p^{j}(\tilde\s)\}=-\delta^{ij}\Delta(\s-\tilde\s).\nn
\end{eqnarray}
The constraints are:
\begin{eqnarray}
&&q^{i} (0)=0,\quad\oint d\s p^{i}(\s)=0,\quad \vec q{\;}'\vec p=0, 
\quad \sqrt{P^{2}}-\oint d\s \sqrt{\vec q{\;}'^{2}+\vec p{\;}^{2}}=0.
\label{conA2}
\end{eqnarray}
Again, the first two constraints belong the second class. 
Calculation of Dirac's brackets on their surface gives expressions (\ref{var2}). 
After that the first two constraints are in involution with all variables; the third constraint defines
the same closed algebra, as $\{x'(\s)p(\s),x'(\tilde\s)p(\tilde\s)\}$; involution of the third
and the fourth constraints can be shown in direct calculation:
\begin{eqnarray}
&&\{q'p,\oint d\tilde\s \sqrt{\tilde q'^{2}+\tilde p^{2}}\}=
\oint d\tilde\s {{q'^{i}\tilde q'^{j}\{p^{i},\tilde q'^{j}\}+p^{i}\tilde p^{j}\{q'^{i},\tilde p^{j}\}
}\over{\sqrt{\tilde q'^{2}+\tilde p^{2}}}}=
-\oint d\tilde\s{{q'\tilde q'+p\tilde p }\over{\sqrt{\tilde q'^{2}+\tilde p^{2}}}}\Delta'(\s-\tilde\s)
\nn\\
&&= -{{d}\over{d\s}} \oint d\tilde\s{{q'\tilde q'+p\tilde p }\over{\sqrt{\tilde q'^{2}+\tilde p^{2}}}}
\Delta(\s-\tilde\s)+\oint d\tilde\s {{q''\tilde q'+p'\tilde p }\over{\sqrt{\tilde q'^{2}+\tilde p^{2}}}}
\Delta(\s-\tilde\s)
=-{{d}\over{d\s}} \sqrt{q'^{2}+p^{2}}+{{q''q'+p'p }
\over{\sqrt{q'^{2}+p^{2}}}}=0.\nn
\end{eqnarray}
Thus, the whole set of constraints (\ref{conA2}) is of the first class.

The third constraint generates the following evolution:
\begin{eqnarray}
&&\delta q(\s)=\{\oint d\tilde\s\tilde\epsilon\cdot(\tilde q'\tilde p),q(\s)\}=\epsilon(\s)q'(\s)-\epsilon(0)q'(0).\nn
\end{eqnarray}
Here the first term corresponds to infinitesimal reparametrization $q(\s)\to q(\s+\epsilon(\s))$,
and the second one -- to a global translation. Because $\delta q(0)=0$, this translation keeps
$q(0)$ in the origin. Variables $p$ are transformed by reparametrizations correctly -- as density:
\begin{eqnarray}
&&\delta p(\s)=\{\oint d\tilde\s\tilde\epsilon\cdot(\tilde q'\tilde p),p(\s)\}=(\epsilon(\s)p(\s))',\nn
\end{eqnarray}
so that cumulative momentum, contained in interval $\s\in[a,b]:\ {\cal P}_{ab}=\int _{a}^{b}d\s p(s)$,
has correct infinitesimal change $\delta {\cal P}_{ab}=\epsilon(\s)p(\s)|_{a}^{b}=
{\cal P}_{b,b+\epsilon(b)}-{\cal P}_{a,a+\epsilon(a)}$, see \fref{f5}.

\begin{center}
\parbox{5cm}{\begin{figure}\label{f5}
\begin{center}
~\epsfysize=3cm\epsfxsize=2cm\epsffile{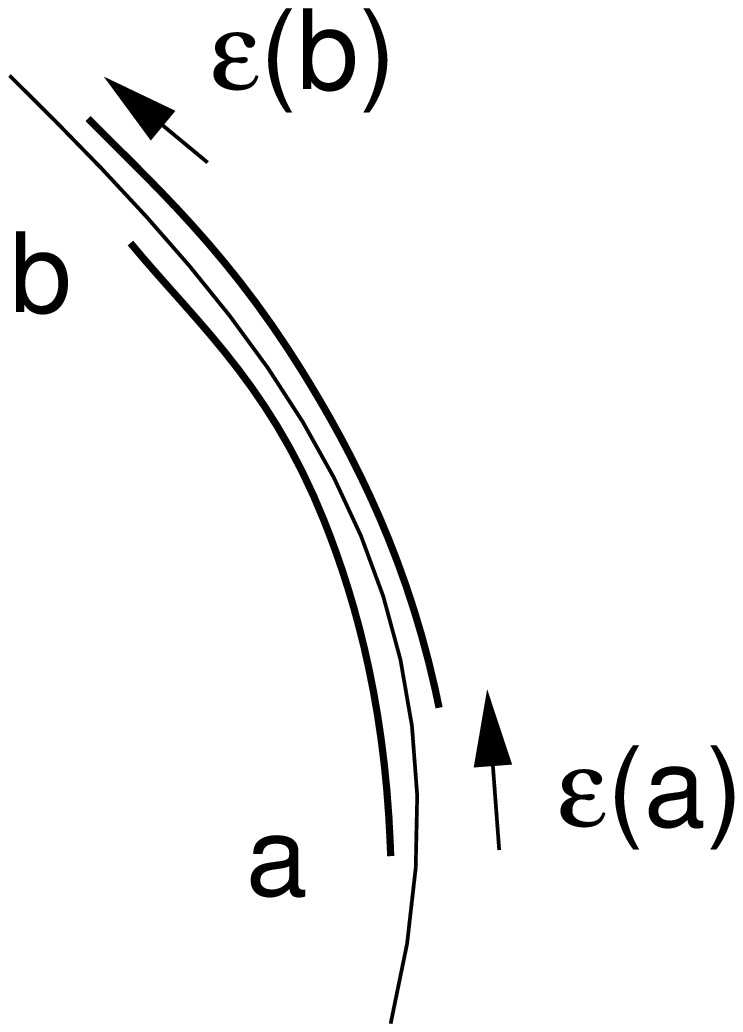}
\end{center}

\end{figure}
}\quad\quad
\parbox{10cm}{\fignum Transformations of segment $[ab]$.}
\end{center}

Let's find the change of variable $x^{i}(0)$. Components $M^{ij}$ are parametrically and translationally
invariant, and the change of $x_{\nu}(0)$ is caused by the second term in (\ref{x0}). The second term
is parametrically invariant and is changed in translation $q^{i}\to q^{i}-\epsilon(0)q'^{i}(0)$ by
$-\epsilon(0)q'^{i}(0)N^{i}_{\nu}$. Thus, the infinitesimal  change of $x^{i}(0)$ equals
$\delta x^{i}(0)=N^{i}_{\nu}
\delta x_{\nu}(0)=\epsilon(0)q'^{i}(0)$: for $x^{i}(\s)=x^{i}(0)+q^{i}(\s)$ 
the translation terms are compensated,
and it is subjected to pure reparametrization.

Evolution, generated by Hamiltonian $H$, is described by equations:
\begin{eqnarray}
&& \dot q^{i} (\s)=\{H,q^{i} (\s)\}= {{p^{i} (\s)}/{p^{0}(\s)}} - 
{{p^{i} (0)}/{p^{0}(0) }} ,\quad
\dot p^{i} (\s)=\{H,p^{i} (\s)\}=({q'^{i} (\s)}/{p^{0}(\s)})',\quad
p^{0}\equiv\sqrt{q'^{2}+p^{2}}.\nn
\end{eqnarray}
It's easy to check, that in such evolution $p^{0}(\s)$ is constant in time: $\dot p^{0}(\s)=0$.
This fact gives a possibility to find from (\ref{x0}) that $\dot x^{i}(0)=p^{i}(0)/p^{0}(0)$,
and reformulate equations in terms of $(x,p)$:
\begin{eqnarray}
&& \dot x^{i} = {{p^{i}}/{p^{0}}},\quad
\dot p^{i}=({x'^{i}}/{p^{0}})'.\nn
\end{eqnarray}
Simplest way to solve these equations is to introduce a special parametrization on the string:
$\hat\sigma=(2\pi/\sqrt{P^{2}})\int_{0}^{\s}d\tilde\s\tilde p^{0}$.
Using the fact, that $x',p$ are transformed under reparametrizations as densities, we have:
\begin{eqnarray}
&& {{dx}\over{d\hat\s}}=x'{{d\s}\over{d\hat\s}}={{\sqrt{P^{2}}}\over{2\pi}} 
{{x'}\over{p^{0}}},\quad
\hat p=p{{d\s}\over{d\hat\s}}={{\sqrt{P^{2}}}\over{2\pi}} 
{{p}\over{p^{0}}},\nn
\end{eqnarray}
and see that in selected parametrization $\hat p^{0}=\sqrt{(dx/d\hat\s)^{2}+\hat p^{2}}$
is constant in $\hat\s$: $\hat p^{0}\equiv\sqrt{P^{2}}/2\pi$. Then
it's possible to rewrite the equations of motion to a simple form\footnote{The condition
$p^{0}=Const$ actually gives Rohrlich's gauge \cite{Rohrlich}. In our case
this condition helps to solve the equations of motion. But being used as a constraint
in the phase space, it leads to complicated redefinition of Poisson brackets,
see the discussion in section \ref{S2}.}:
\begin{eqnarray}
&& {{dx}\over{d\tau}} ={ {2\pi} \over {\sqrt{P^{2}}} } \hat p,\quad 
{{d\hat p}\over{d\tau}} = { {2\pi} \over {\sqrt{P^{2}}} } 
{{d^{2}x}\over{d\hat\s^{2}}}\ \Rightarrow\
{{d^{2}x}\over{d\tau^{2}}} ={ {(2\pi)^{2}} \over {P^{2}} } 
{{d^{2}x}\over{d\hat\s^{2}}}.\nn
\end{eqnarray}
Solving the obtained 2-dimensional wave equation, we have
\begin{eqnarray}
&& x= f\left({ {2\pi} \over {\sqrt{P^{2}}} } \tau+\hat\s\right)+
g\left({ {2\pi} \over {\sqrt{P^{2}}} } \tau-\hat\s\right),\nn\\
&&\hat p= f'\left({ {2\pi} \over {\sqrt{P^{2}}} } \tau+\hat\s\right)+
g'\left({ {2\pi} \over {\sqrt{P^{2}}} } \tau-\hat\s\right),\nn
\end{eqnarray}
where $f,g$ are $2\pi$-periodical functions. From the constraints $x'p=0,\ \hat p^{0}=\sqrt{P^{2}}/2\pi$
we also have $f'^{2}=g'^{2}=P^{2}/(4\pi)^{2}$. Finally, we see that the resolved evolution has
period $\Delta\tau=\sqrt{P^{2}}$.

\vspace{2mm}
\noindent{\it Note:} periodicity of string dynamics has been established by many methods.
The work \cite{sing} gives purely geometrical explanation of this property.

\vspace{2mm}
\noindent 3. To obtain the expression (\ref{Lor}) for Lorentz generators, we write:
$$M_{\mu\nu}= N_{\mu}^{\alpha}N_{\nu}^{\beta}M^{\alpha\beta}=
- N_{[\mu}^{i}N_{\nu]}^{0} M^{i0}+N_{\mu}^{i}N_{\nu}^{j}M^{ij}.$$
Then, using (\ref{Mi0}), we rewrite the first term as
$$-x^{i}(0)N_{[\mu}^{i}P_{\nu]}-N_{[\mu}^{i}N_{\nu]}^{0} \oint d\s q^{[i}p^{0]}=
\left(x_{\mu}(0)-N_{\mu}^{i}{{1}\over{\sqrt{P^{2}}}}\oint d\s q^{[i}p^{0]}\right)P_{\nu}
-(\mu\leftrightarrow\nu)$$
and using (\ref{x0}):
$$=\left(Z_{\mu}-\half N^{i}_{\rho}(\df N^{j}_{\rho}/\df P_{\mu})M^{ij}\right)
P_{\nu}
-(\mu\leftrightarrow\nu).$$
Then, defining $X_{\mu}=Z_{\mu}-\half N^{i}_{\rho}(\df N^{j}_{\rho}/\df P_{\mu})M^{ij}$,
we obtain (\ref{Lor}).

\vspace{2mm}
\noindent{\it Note:} mean coordinate $Z_{\mu}$ is not Lorentz vector,
but has more complicated law of transformation \cite{slstring}.
The reason is that variables $N_{\mu}^{i}$, {\it being functions
of $P_{\mu}$ only}, cannot be Lorentz vectors\footnote{They should be transformed by 
a subgroup of Lorentz transformations, not changing $P_{\mu}$, and this contradicts
to the fact, that they are functions of $P_{\mu}$.}. However, variables $M_{\mu\nu}$,
composed from these Lorentz non-covariant objects, define Lorentz tensor and generate correct
Lorentz algebra both on classical and quantum levels. Proof of this fact can be found in
\cite{slstring}.

\section*{Appendix 2: commutators of H with constraints}
The first constraint in (\ref{conq2}) is satisfied identically and commutes with all operators.
The second constraint commutes with operator $\vec p$, defined by (\ref{varq3}),
and as a result, with each term in expansion (\ref{H}). Let's show that the third constraint commutes
with each term in (\ref{H}). From the explicit definition of operator $\vec p$ it's easy to obtain
the following commutation relations:

\begin{eqnarray}
&& 
[q'p,\tilde p]=i\Delta'(\s-\tilde\s)p,\quad
[q'p,\tilde q']=i\Delta'(\s-\tilde\s)q',
\nn\\
&& 
[q'p,\tilde p^{2}]=2i\Delta'(\s-\tilde\s) (p\tilde p),\quad
[q'p,\tilde q'^{2}]=2i\Delta'(\s-\tilde\s)(q'\tilde q'),
\nn\\
&&
[q'p, (\tilde p^{2})^{n} ]=2in\Delta'(\s-\tilde\s) (p\tilde p) 
(\tilde p^{2})^{n-1} ,\quad
[q'p, (\tilde q'^{2})^{-n+1/2} ]=2i(-n+1/2)\Delta'(\s-\tilde\s)(q'\tilde q') 
(\tilde q'^{2})^{-n-1/2} ,
\nn\\
&&
[q'p, (\tilde q'^{2})^{-n+1/2} (\tilde p^{2})^{n} ]=2i\Delta'(\s-\tilde\s) 
( (-n+1/2)(q'\tilde q')
(\tilde q'^{2})^{-n-1/2} (\tilde p^{2})^{n} +n(\tilde q'^{2})^{-n+1/2} 
(p\tilde p) (\tilde p^{2})^{n-1} ) =
\nn\\
&&
=2i\left(
{{d}\over{d\s}}[\Delta(\s-\tilde\s) ( (-n+1/2)(q'\tilde q')
(\tilde q'^{2})^{-n-1/2} (\tilde p^{2})^{n} +n(\tilde q'^{2})^{-n+1/2} 
(p\tilde p) (\tilde p^{2})^{n-1} )] \right.\nn\\
&&\left.-\Delta(\s-\tilde\s) [ (-n+1/2)(q''\tilde q')
(\tilde q'^{2})^{-n-1/2} (\tilde p^{2})^{n} +n(\tilde q'^{2})^{-n+1/2} 
(p'\tilde p) (\tilde p^{2})^{n-1} ]
\right).
\nn
\end{eqnarray}
Integrating the result by $\oint d\tilde\s$, we have
\begin{eqnarray}
&&2i\left({{d}\over{d\s}}[1/2\cdot (q'^{2})^{-n+1/2} (p^{2})^{n} ]-
[(-n+1/2)(q''q') (q'^{2})^{-n-1/2} (p^{2})^{n} +n(q'^{2})^{-n+1/2} (p'p) (p^{2})^{n-1} ]\right)=0.\nn
\end{eqnarray}

\end{document}